\documentclass{mscs}
\usepackage{alltt}
\usepackage{amsmath,amsfonts,amssymb,amstext,eucal}
\usepackage{url}
\title{Affine functions and series with co-inductive real numbers}
\newcommand{\coq}{{\sc Coq}}
\author{Yves Bertot}
\date{November 2005}
\begin{document}
\maketitle
\begin{abstract}
We extend the work of A. Ciaffaglione and P. di Gianantonio on
mechanical verification of algorithms for exact computation on real numbers,
using infinite streams of digits implemented as a co-inductive type.
Four aspects are studied: the first aspect concerns the proof
that digit streams can be related to axiomatized real numbers when they
are already present in the proof system.  The second aspect re-visits
the definition of an addition function, looking at techniques to let the
proof search engine
perform the effective construction of an algorithm that is correct by
construction.  The third aspect concerns the definition of a
function to compute affine formulas with positive rational
coefficients.  This is an example where we need to
 combine co-recursion and recursion.  The fourth aspect concerns the
definition of a function to compute series, with an application on the
series that is used to compute Euler's number \(e\).  All these
experiments should be reproducible in any proof system that supports
co-inductive types, co-recursion and general forms of terminating
recursion.  We used the \coq{} system \cite{coq,coqart,Gimenez94}.
\end{abstract}
\section{Introduction}
Several proof systems provide data-types to describe real numbers,
together with basic operations and theorems giving 
 an ordered, complete, and archimedian field
\cite{HOL-light,HarrisonThesis, mayero-thesis}.  In the \coq{} system, several
approaches have been taken; depending on whether developers wanted to
adhere to pure constructive mathematics or more classical approaches.
In the classical approach, the type of real numbers is merely
``axiomatized'', the existence of the type and the elementary
operations is assumed and the properties of these operations are
asserted as axioms.  This approach has been used extensively to
provide a large collection of results, going all the way to the
description of trigonometric functions, calculus and the like.
However, because the type of real numbers is axiomatized, there is no
``physical representation of numbers'' and the basic operations
correspond to no algorithm.  

In an alternative approach, a type of {\em constructive} numbers may be
defined as a data-type and the basic operations may
be described as algorithms manipulating elements of this data-type.
This approach is used for instance in C-CoRN 
\cite{DBLP:conf/mkm/Cruz-FilipeGW04}.  
A.~Ciaffaglione and P.~di~Gianantonio \cite{ciaffaglione.00} showed that
a well-known representation of real numbers as infinite sequences
of redundant digits could easily be implemented inside theorem provers
with co-inductive types.  We say the digits are redundant
because several representation are possible for every number.  In the
case of \cite{ciaffaglione.00} the representation is simply inspired
by the usual binary representation of fractional numbers and
made more redundant by adding the possibility to use a negative digit.
A.~Ciaffaglione and P.~di~Gianantonio then provide addition,
multiplication, and show that these operations enjoy the properties
that are expected from a constructive field of real numbers.

In our approach there is also an extra digit, but its meaning is
intermediary between the existing 0 and 1.  Edalat and Potts \cite{edalat.98}
show that both approaches are special cases of a more general family of
representations based on {\em linear fractional transformations}.

Once we have chosen the way to represent real numbers as a data-type,
we proceed by establishing a relation between this data-type and the
axiomatized type of real numbers.  This is a departure from the
prescription of pure constructive mathematics, because we rely on the
non-constructive axioms of that theory to state the correctness of our
algorithms.  Still, we also describe the relation between our representation
and constructive approaches to real numbers, by showing how infinite
sequences of digits can be produced from constructive views of Dedekind
cuts or Cauchy sequences.

Once we have provided the basic data-type and its relation with the
axiomatized theory of real numbers, we proceed by defining an addition
function.  We rely on proof search tools to 
construct the addition algorithm: we only provide
guidelines for the construction of the algorithm, without actually describing
all 25 cases in the function.  The correctness proof then consists
in showing that there is a morphism between the data-type and the
axiomatized type.  Our contribution in this part is to show how to
use the proof search engine to construct a well formed addition function.

We then focus on affine formulas combining two
real values with rational coefficients.  For these more general
operations, we need to combine co-recursion and well-founded
recursion. We show that the function responsible for producing the
infinite stream of digits representing the result can be decomposed
in two recursive functions.  One of the functions is a guarded
co-recursive function as proposed in \cite{Gimenez94}, the second
function is a well-founded recursive function as in \cite{nordstrom88}.  Each
function satisfies a different form of constraint: the co-recursive
function does not need to be terminating, but it must produce at least
a digit at each recursive call, while the well-founded function does
not need to produce a digit at each recursive call, but it must
terminate.  Our main contribution in this part is to show that some
functions that appear at first sight to be outside the expressive
power of guarded co-recursion can actually be modeled and proved
correct.

In a fourth part, we focus on constructively converging infinite sums.
  We show how to avoid having to consider the infinity of
terms that are parts of the sum.  We exhibit a
framework that can be re-used from one series to the other.  As
applications, we show how to compute the infinite stream representing
Euler's number \(e\) and to multiply two real numbers represented as
infinite streams of digits.  In particular, the algorithm we obtain can be
executed directly using the reduction mechanism provided in \coq{} to
compute \(e\) to a great precision in a reasonable time.

Our account stops here, although the experiments described in this paper
seem to open the door for a more complete study of real functions,
especially analytic functions.

\section{Related work}
For numerical computation, real numbers are usually represented as
approximates using floating point numbers.  These floating point
numbers are composed of a mantissa and an exponent, so that the
value of the least significant bit in the mantissa varies with the
exponent.  Still both the exponent and the mantissa have a fixed size, 
so that there is only a finite number of floating point numbers and real
values must be rounded to find the closest floating point numbers.
Floating-point based computations are thus only approximative and
errors stemming from successive rounding operations may accumulate to
the point that some computations can become grossly wrong
\cite{JMMuller}.

In spite of their limitations, Floating-point numbers are used
extensively: most processors
directly provide an implementation of the elementary operations
(addition, subtraction, multiplication, division) according to a
standard that gives a precise mathematical meaning to the rounding
operations \cite{IEEE754}.  This standard provides the basis to
implement computations with a guaranteed precision
\cite{JMMuller,MPFR}, sometimes with correctness proofs that can be
verified with the help of computer-aided proof tools
\cite{DauRidThe01, tphols2000-Harrison,Russinoff99,
Boldo04}.  In particular, some approaches, named {\em expansions} make
it possible to increase drastically the number of representable numbers
by extending the length of the mantissa \cite{BolDauMorThe02}.

An alternative to floating point and rounding concentrates on
data-structures that support exact computation.
  Among the possible approaches, the best well-known are based
on continued fractions \cite{gosper.72, vuillemin.90} or
representations with mantissas that grow as needed
\cite{MenissierReals95}.  In the latter case, the representation is
very close to the floating-point representations with rounding modes.
One way to understand this ``growing mantissa'' is to view it
as an infinite sequence of digits, where only a finite prefix is
known at any time.  An introduction to exact real arithmetics
can be found in \cite{edalat.98}.
Implementations are provided in the setting of conventional programming
languages {\cite{DBLP:conf/cca/Lambov05,DBLP:conf/cca/Muller05}}, 
or in the setting of functional programming
{\cite{DBLP:conf/lfp/BoehmCRO86,menissier.94,BauerEscardoSimpson02}}.

Formal proofs about computations on infinite data-structures are a
privileged ground for the use of co-inductive types
\cite{CoquandInfinite93,Gimenez94}.  First experiments on the topic
of exact real number computations using co-inductive types were
performed by Ciaffaglione and di~Gianantonio {\cite{ciaffaglione.00}}
who showed that one could represent infinite sequences of digits with
co-inductive types and the basic operations of arithmetics (addition,
multiplication, comparison) with simple co-recursive functions, as
long as the set of digits was extended to allow for enough redundancy.
Niqui \cite{NiquiThesis} also studies the problems of modeling real
number arithmetic for use in formal proofs, providing a single point
of view to account for both continued fractions and infinite sequences of
digits.  Our approach is very similar to Ciaffaglione and 
di~Gianantonio's, it only differs in the collection of digits that we
consider.

The example of combination of co-recursive functions and well-founded
recursive functions that we exhibit in our treatment of affine
formulas is related to work by di Gianantonio and Miculan
\cite{diGianantonioMiculan} and our own work on partial co-recursive
functions \cite{bertot05a}.

Concerning the computation of series, we are aware through personal
communication that computations of \(e\) had already been done using the
real numbers as they are formalized in the C-CoRN library
\cite{DBLP:conf/mkm/Cruz-FilipeGW04}; it seems that co-inductive presentations
make it possible to achieve a higher efficiency.  While the C-CoRN library
aims at providing a comprehensive study of constructive mathematics, our
work abandons some of the foundations of constructive mathematics:
we let our logical
reasoning depend on non-constructive arguments, although we do provide
algorithms working on concrete data-structure to represent real numbers and
the basic operations.  We believe the algorithms we develop will be useful
even in the context of constructive mathematics, but the proofs of correctness
will probably have to be re-done.

\section{Redundant digit representation for real numbers}
We are all used to the notation with a decimal point to represent
real numbers.  For instance, we usually write a number between 0 and 1
as a string of the form \(0.1354647\cdots\) and we know that the
sequence of digits must be infinite for some numbers, actually all
those that are not of the form \(\frac{a}{10^b}\), where \(a\) and
\(b\) are positive integers.  It is a bit less natural, but still easy
to understand, that all numbers can be represented by an infinite
sequence: for those that have a finite representation, it suffices to
add an infinite sequence of zeros.  Moreover the number 1 can also be
represented by the sequence \(0.999\cdots\)

When we know a prefix of one of these infinite sequences, we actually
know the number that is represented up to a certain precision.  If the
prefix has length \(p\), we actually know precisely the bounds of an interval
of width \(\frac{1}{10^p}\) that contains the number.  We are accustomed
to reasoning with these prefixes of infinite sequences and we expect
tools to return correct prefixes of an operation's output when this
operation has been fed with correct prefixes for the inputs.

In the conventional representation, the number {\emph{ten}}
plays a special role:
it is the {\em base}.  We can change the base and use digits that are
between 0 and the base.  For instance, we can use {\emph{two}} as the base, so
that the digits are only 0 and 1.  The number \(\frac{1}{2}\) can then
be represented by the sequence \(0.1000\cdots\) and the number 1 can
be represented by the sequence \(0.1111\cdots\).  For a sequence
\(0.d_1d_2\cdots\), the number being represented is:
\[\sum_{i=1}^\infty \frac{d_i}{2^i}.\]
the following equalities hold:
\begin{eqnarray*}
  0.0s&=&\frac{0.s}{2}\\
  0.1s&=&\frac{0.s+1}{2}
\end{eqnarray*}
A prefix with \(p\) digits gives
an interval of width \(\frac{1}{2^p}\) that contains the number
represented by the infinite sequence.  In the rest of this paper, we
will carry on using base 2 (but the set of
digits will change).

For the computation of basic operations, the base-2 conventional
representation is not really well adapted.  Here is an example that
exhibits the main flaw of this representation.  The numbers
\(\frac{1}{3}\) and \(\frac{1}{6}\) add up to give \(\frac{1}{2}\).
However, the infinite sequences for these numbers are given in the
following equations:
\begin{eqnarray*}
 \frac{1}{3}&=&0.01010101\cdots\\
\frac{1}{6}&=&0.00101010\cdots\\
\frac{1}{2}&=&0.10000000\cdots = 0.01111111\cdots
\end{eqnarray*}
The following reasoning steps justify the first equation:
\[
0.01010101\cdots=\sum_{i=1}^\infty \frac{1}{2^{2i}}
                =\frac{1}{1-\frac{1}{4}}-1=\frac{1}{3}\]

Similar justifications can be used for the other equations.
If \(w\) is a prefix of \(0.0101\cdots\), then \(w\) is also the prefix
of all numbers between \(w00\cdots\) and \(w111\cdots\).  These two
numbers are rational numbers of the form \(\frac{a}{2^b}\) and neither
can be equal to \(\frac{1}{3}\).  Thus, we actually have
an interval of possible values that contains both values that are
smaller and values that are larger than \(\frac{1}{3}\).  The same
property occurs for the representation of \(\frac{1}{6}\).  When
considering the sum of values in the interval around \(\frac{1}{3}\)
and values in the interval around \(\frac{1}{6}\), the results are in
an interval that contains both values that are smaller and values that
are larger than \(\frac{1}{2}\).  However, numbers of the form
\(0.1\cdots\) can only be larger than or equal to \(\frac{1}{2}\) and
numbers of the form \(0.0\cdots\) can only be smaller than or equal to
\(\frac{1}{2}\).
Thus, even if we know the inputs with a great precision, we must
indefinitely delay the decision and require more precision on the
input before choosing the first digit of the result: we need to know
the inputs with infinite precision before deciding the first digit of
the output.

We solve this problem by adding a third digit in the notation.  This
digit provides a way to express that the interval given by a prefix
has \(\frac{1}{2}\) in its interior.  This new digit adds more
redundancy in the representation.  We now have three digits, even
though we still work in base 2.  We choose to name the three
digits  {\tt L} (for {\em left}), {\tt R} (for {\em Right}), and
{\tt C} (for {\em center}).
\begin{itemize}
  \item The digit {\tt L} is used like the digit \(0\).  If \(x\) is
an infinite sequence of digits representing the number \(v\), the
sequence \({\tt L}x\) represents \(v/2\).
\item The digit {\tt R} is used like the digit \(1\).  If \(x\) is
an infinite sequence of digits representing the number \(v\), the
sequence \({\tt R}x\) represents \(v/2+1/2\).
\item The digit {\tt C} is used with the following meaning:  if \(x\) is
an infinite sequence of digits representing the number \(v\), the
sequence \({\tt C}x\) represents \(v/2+1/4\).
\end{itemize}
The fact that {\tt L} is like 0, {\tt R} is like 1, etc can also be
expressed using a function \(\alpha\) such that \(\alpha({\tt L})=0\),
\(\alpha({\tt R})=1\) and \(\alpha({\tt C})=\frac{1}{2}\).  With this
function we can still interpret a digit sequence \(0.d_1d_2\cdots\) as
an an infinite sum:
\[\sum_{i=1}^\infty \frac{\alpha(d_i)}{2^i}\]

From now on, we consider only numbers in the interval \([0,1]\) and we
drop the first characters ``0.'' when writing a sequence of digits.
We use the same notation for a sequence of digits and the
real number it
represents.  In the same spirit, we use the same notation for a
digit \(d\) and the function it represents, the function that maps \(x\) to
\(\frac{x+\alpha(d)}{2}\).  Last, we 
associate the digits {\tt L}, {\tt R}, {\tt C} to the intervals
\([0,\frac{1}{2}]\), \([\frac{1}{2},1]\), and
\([\frac{1}{4},\frac{3}{4}]\), respectively.  These interval are
called {\em basic intervals}.

The redundancy of the new digit gives a very simple property:
a number that can be written \({\tt
  CL}x\) can also be written \({\tt LR}x\) and a number that
can be written \({\tt CR}x\) can also be written \({\tt RL}x\).  This
property is used several times in this paper.

\subsection{Formal details of infinite sequences and real numbers}
In our formalization, we benefit from theories that state the main
properties of natural numbers (type {\tt nat}), integers (type {\tt
Z}), and real numbers (the type is usually written {\tt R}, but in
this article we shall write it as {\tt Rdefinitions.R} to avoid
ambiguity with the ``digit'' {\tt R}).  The two integer types come
with addition, subtraction, and multiplication, while the type of real
numbers is also equipped with division.  The integer types are actually
described as inductive types and the basic operations are implemented
as recursive functions.  For the real numbers, the existence of the
type, two constants 0 and 1, the operations, comparison predicates,
and the properties of these operations (associativity, distributivity,
inverse, etc.) are assumed.  Among the assumed features, there is an
axiom that expresses completeness, which states that every bound and non-empty
subset of \(\mathbb{R}\) has a least upper bound in \(\mathbb{R}\).
This means that whenever we exhibit a property \(E\)
and prove that it is bound, we can construct a function that returns
its least upper bound.  This completeness axiom is inherently non-constructive.
To be more precise, our work describes a collection of algorithms on
a representation of real numbers, in some sense a constructive of 
real numbers, but many justifications of correctness, which rely on
the axiomatized real numbers, are non-constructive.

The axiomatization
of real numbers also provides a few decision procedures.  The decision
procedure {\tt field} \cite{del.may.2001} solves equalities between
rational expressions, occasionally leaving proof obligations to make
sure denominators are non-zero.  The decision procedure {\tt fourier}
determines when a collection of in-equations concerning affine formulas
with rational coefficients is satisfiable.

The type of digits is described as an enumerated type:
\begin{alltt}
Inductive idigit: Set := L | R | C.
\end{alltt}
We provide both a {\em numeric} interpretation (named {\tt alpha}) and a 
{\em functional} interpretation (named {\tt lift}) to these digits.  These
can be defined in {\coq} with the following text:

\begin{alltt}
Definition alpha (d:idigit): Rdefinitions.R :=
  match d with L => 0 | C => 1/2 | R => 1 end.

Definition lift (d:idigit)(x:Rdefinitions.R) := (x+ alpha d)/2.
\end{alltt}
The type of infinite sequences of digits is based on a polymorphic
type of streams, which is defined as a co-inductive type using a declaration
with the following form:
\begin{alltt}
CoInductive stream (A:Set): Set := Cons: A -> stream A -> stream A.
\end{alltt}
This definition defines {\tt stream A} to be a data-type for any type
{\tt A}.  It also defines the constructor {\tt Cons}, with the type
given in the definition.  This definition is similar to a recursive
data-type definition in a conventional functional programming language.
In our mathematical notation, we will simply
write \(ds\) instead of {\tt Cons \(d\) \(s\)}.  In {\coq} excerpts,
we will also use the notation \(d{\tt ::}s\).

In proof systems, recursion is seldom unrestricted.  In the {\coq} system,
it is mostly provided as a companion to inductive and co-inductive
data-structures.  For inductive structures, the form of recursion that
is provided is called {\em structural} recursion, and it basically imposes
that a recursive function takes an element of inductive type as argument
and a recursive call can only be performed if the argument is
a sub-term of the initial argument.  For co-inductive structures, the form
of recursion that is provided is called {\em guarded co-recursion}, and
it basically imposes that a co-recursive value must be
an element of a co-inductive type and that co-recursive calls can only be used
to produce sub-terms of the output.  More general forms of recursion are
also provided, for instance {\em well-founded recursion}, where recursive
calls are allowed only if the argument of the recursive call is a predecessor
of the initial argument with respect to a relation that is known to be
well-founded (which intuitively means that this function contains no infinite
chain of predecessors).  Well-founded recursion can actually be shown to
be a special case of structural recursion \cite{nordstrom88,moh93,coqart}.

Co-recursive values need not be functions.  For instance, 0 and 1 are
represented by the infinite sequences {\tt LLL\dots} and {\tt
RRR\dots}, these are defined as co-recursive values with the following
definition:
\begin{verbatim}
CoFixpoint zero: stream idigit := L::zero.

CoFixpoint one: stream idigit := R::one.
\end{verbatim}
To relate infinite streams of digits with real numbers 
we define a co-inductive relation.
\begin{verbatim}
CoInductive represents: stream idigit -> Rdefinitions.R -> Prop:=
  repr: forall d s r,
    represents s r -> 0 <= r <= 1 -> represents (d::s) (lift d r).
\end{verbatim}
This definition introduces both the two-place predicate {\tt represents}.
and a constructor, named {\tt repr}, which can be used as a theorem to
prove instances of this predicate.  The statement of this theorem can
be read as ``for every {\tt s} and {\tt r}, if the proposition
{\tt represents s r} holds and the proposition {\tt 0 <= r <= 1} hold,
then the proposition {\tt represents (d::s) (lift d r)} holds''.
This relation really states that infinite streams are only used to 
represent numbers between 0 and 1 and it confirms the correspondence
between the digits and their function interpretations.

An alternative approach to relating sequences of digits and real numbers
is to build a function that maps an infinite sequence to a real value.
Every prefix of an infinite sequence corresponds to an interval that
contains all the values that could have the same prefix.
As the prefix grows, the new intervals are included in each other,
and the size is divided by 2 at each step, while the bounds remain rational.
We defined a function {\tt
  bounds} to compute the interval corresponding to the prefix of a
given length for a given sequence.  This function takes as
arguments a digit sequence and a number \(n\) and it computes the
bounds of an interval that contains all the real numbers whose
representation shares the same prefix of length \(n\).  This function
is primitive recursive in \(n\) (in other words, it is structural recursive
with respect to the conventional representation of natural numbers as
an inductive type).
\begin{eqnarray*}
  {\tt bounds}(\ldots, 0)&=&[0,1]\\
  {\tt bounds}(ds, n+1)&=&[{\tt lift}~d~a,{\tt lift}~d~b]\quad\hbox{where}\quad{\tt  bounds}(s, n)=[a,b]
\end{eqnarray*}
In practice, we do not manipulate real numbers in this function, but only
integers.  The result of the function is a structure \(((a,b),k)\) such
that the interval is \([\frac{a}{2^k},\frac{b}{2^k}]\).

We then define a function that maps a stream of digits to a sequence
of real numbers, which are the lower bounds of the intervals.
This function is called {\tt si\_un} and is defined by the following
text, where {\tt IZR} is the function that injects integers
in the type of real numbers.
\begin{verbatim}
Definition si_un (s:stream idigit) (n:nat): Rdefinitions.R :=
   let (p,k) := bounds n s in let (a,b) := p in IZR a/IZR(2^k).
\end{verbatim}
We prove that for every \(d\), {\tt lift \(d\)} is monotonic, so that
{\tt si\_un \(s\)} is a growing sequence bounded by 1.
All this leads us to a proof that {\tt si\_un \(s\)} has a limit and
that this limit is in [0,1].  This makes it possible to define
the function {\tt real\_value} that associates an infinite sequence
of digits to the limit.  

We then prove that adding a digit in front of a sequence is the
same as using this digit as a function.
\begin{verbatim}
 Theorem real_value_lift:
    forall d s, real_value (d::s) = lift d (real_value s).
\end{verbatim}
This makes it possible to show that {\tt real\_value} and {\tt represents}
follow the same structure and to obtain the following theorem.
\begin{verbatim}
Theorem represents_real_value: forall s, represents s (real_value s).
\end{verbatim}
To complete the correspondence between the two notions we need to
express that the relation {\tt represents} is actually a function.
We do this by expressing that the distance between two possible
values represented by a sequence is smaller than \(\frac{1}{2^n}\):
This is proved by induction over \(n\).
\begin{verbatim}
Theorem represent_diff_2pow_n :
  forall n x r1 r2, represents x r1 -> represents x r2 ->
        -1/(2^n) <= r1 - r2 <= 1/(2^n).
\end{verbatim}
It is then easy to conclude with the following theorem.
\begin{verbatim}
Theorem represents_equal: forall s r, represents s r -> real_value s = r.
\end{verbatim}
We thus have two ways to express that a given sequence represents a
real number.  The function {\tt real\_value} is more pleasant to use
in theorem statements, but the co-inductive property makes proofs more
elegant.  Actually, all proofs of function correctness presented in
this paper are performed using a proof by co-induction based on
{\tt represents}, even though theorem statements are sometimes
expressed using {\tt real\_value}.

This raises a third question: {\em given an arbitrary real number, is
there a digit stream that represents it?}  The answer to this question is
related to question of constructivism in mathematics.  If you have
a constructive description of your real number, then this may for instance
be tantamount
to a boolean predicate on rational numbers corresponding to this
number viewed as a Dedekind cut (a boolean predicate that is {\tt false}
for every rational number smaller than represented real number and
{\tt true} for any rational number that is larger).
We can produce the co-recursive value
corresponding to any boolean predicate using a co-recursive function.  Here
is an example, where the rational numbers are viewed as pairs of
integers ({\tt p (\(a\), \(b\)) = true} means that the real number of interest
is smaller than or equal to \(\frac{a}{b}\)):
\begin{alltt}
CoFixpoint stream_of_cut (p:Z*Z->bool) : stream idigit :=
  match p (1, 2) with
    true => R::stream_of_cut (fun r => let (a,b) := r in p (a+b, 2*b))
  | false => L::stream_of_cut (fun r => let (a,b) := r in p (a, 2*b))
  end.
\end{alltt}
Alternatively, the real number of interest can be given by a Cauchy sequence
of rational numbers and a constructive proof that it satisfies the Cauchy
criterion.  One way to describe Cauchy sequences is to fix a function
\(h\) from {\tt nat} to \(\mathbb{R}\), with its limit in 0 when the argument
goes to infinity.  A Cauchy sequence may then be given by a function \(f\)
from {\tt Z} to \(\mathbb{Q}\) and the Cauchy criterion may be given
by a monotonic function \(g\) from {\tt Z} to {\tt Z} such that
\[\forall n~m~p, g(n)\leq m \wedge g(n)\leq p \Rightarrow |f(m)-f(p)| < h(n)\]
To construct the infinite list of digits for a given Cauchy sequence, we
simply need to repeat the following process:
\begin{enumerate}
\item compute the \(n\) first elements of the stream, this actually gives
an interval of width \(\frac{1}{2^n}\), compute the lower bound
\(b\) of this interval,
\item find the least \(n\) such that \(h(n)\leq \frac{1}{2^{n+3}}\),
\item compute \(f(g(n))\), we know that the distance between this value and
the sequence's limit is less than \(\frac{1}{2^{n+3}}\),
\item compute the value \(a=2^n(f(g(n))-b)\), this value is in [0,1]
by an invariant of the recursive process,
\item if \(a\leq \frac{3}{8}\), then we know that for every \(m\geq g(n)\),
\(f(m) \leq \frac{1}{2}\) and we choose the \(n+1\)th digit to be {\tt L},
\item if \(\frac{3}{8}\leq a \leq \frac{5}{8}\), then we know that for
every \(m\geq g(n)\), \(\frac{1}{4}\leq f(m)\leq \frac{3}{4}\), and we
choose the \(n+1\)th digit to be {\tt C},
\item if \(\frac{5}{8}\leq a\), then we know that for every \(m\geq g(n)\),
\(\frac{1}{2}\leq f(m)\), and we choose the {n+1}th digit to be
{\tt R}
\end{enumerate}
We contend that this technique gives a constructive process to associate
a sequence of digits to a sequence of rational numbers associated with
a constructive proof that this sequence satisfies the Cauchy criterion.

When implementing this recursive process as a co-inductive function,
we propose to represent rational numbers with pairs of integers and
to replace comparisons of rational numbers with comparisons of integers.
It is also more convenient to restrict ourselves to the case where 
\(h(n)=\frac{1}{2^n}\).  In the recursive process, we do not need to
keep the list of the \(n\) first digits, we only need to know \(n\) and
the lower bound of the represented interval, these are given as arguments
to the co-recursive function:
\begin{alltt}
CoFixpoint stream_of_cauchy (f: Z -> Z*Z)(g: Z -> Z)
  (n:Z)(b:Z*Z) : stream idigit :=
  let (vn,vd) := f(g n) in
  let (bn,bd) := b in
  let (d, r) := 
    if is_smaller (8*2^n*(vn*bd-vd*bn)) (3*vd*bd) then
       (L,b)
    else if is_smaller (8*2^n*(vn*bd-vd*bn)) (5*vd*bd) then
       (C,(4*2^n*bn+bd,4*2^n*bd))
    else (R, (2*2^n*bn+bd,2*2^n*bd)) in
  let (new_bn, new_bd) := r in
     d::stream_of_cauchy f g (n+1) (new_bn, new_bd).
\end{alltt}
The functions {\tt stream\_of\_cut} and {\tt stream\_of\_cauchy} are
only given here to show the feasibility of connecting streams of
digits with the usual notions of real number constructions,
but more efficient ways to handle rational numbers should
be used if these functions were to be used effectively, for instance
least common denominators should be computed between fraction numerators
and denominators.

For an arbitrary real number between 0 and 1 given non constructively,
but known by its binary representation (an infinite sequence of 0 and 1
digits), this real number is simply represented by the corresponding
infinite stream where 0 is replaced with {\tt L} and 1 is replaced with
{\tt R}; however, the fact that the real number is given non constructively
is reflected by the fact that we can't write a co-recursive function that
produces the stream.

\subsection{Addition}
It is well-known\footnote{P. Martin-L{\"o}f suggested to the author
that Cauchy had already devised an algorithm for addition in a similar
representation.  Di Gianantonio refers to Cauchy and Leslie, but the
reference to Cauchy's work is wrong and the reference to Leslie could not
found at the time of writing.} that adding
 two infinite sequences of redundant digits
can be described as a simple automaton that reads digits from both
inputs and produces digits at every recursive call.  Two approaches
can be taken: either this automaton is understood as a program that
keeps a carry as it processes the inputs, or it can be viewed as a
program that performs a little look-ahead before outputting the result
and processing the rest, maybe with a slight modification of the first
digit in both inputs.  The algorithm we describe follows the second
approach.

Our algorithm has two parts.  The first part is a function that
computes the arithmetic mean of two values (in other words, the
half-sum).  The second part is a function that computes the double
of a value.  The first algorithm maps two real numbers in \([0,1]\) to 
a value in \([0,1]\).  The doubling function only returns a meaningful
value when the input is smaller than or equal to \(\frac{1}{2}\).

For the half-sum, the structure of the algorithm is as follows: if the
inputs have the form \(d_1d_2x\) and \(d_3d_4y\), then the algorithm
outputs a digit \(d\) and calls itself recursively with the new arguments
\(d_5x\) and \(d_6y\).  Written as an equation, this yields the following
formula:
\[\hbox{\tt half\_sum}(d_1d_2x,d_3d_4y)=d({\tt half\_sum}(d_5x,d_6y)).\]

Here is an example, suppose that \(d_1={\tt L}\) and \(d_3={\tt R}\),
in this case we can choose \(d={\tt C}\) and \(d_5=d_2\) and
\(d_6=d_4\), because the following equalities hold, using the
interpretations of digits as functions:
\begin{eqnarray*}
\hbox{\tt half\_sum}({\tt L}d_2x,{\tt R}d_4y)&=&
\frac{  {\tt L}d_2x+{\tt R}d_4y}{2}\\
&=&\frac{
\frac{d_2x}{2}+\frac{d_4y}{2}+\frac{1}{2}}
{2}\\
&=&\frac{d_2x}{4}+\frac{d_4y}{4}+\frac{1}{4}\\
&=&\frac{\frac{d_2x+d_4y}{2}}{2}+\frac{1}{4}\\
&=&{\tt C}(\hbox{\tt half\_sum}(d_2x,d_4y)).
\end{eqnarray*}
In this case, it is not necessary to scrutinize \(d_2\) and \(d_4\) to
decide the value of \(d\) and the arguments for the recursive call.
The equation can be re-written as
\[{\tt half\_sum}({\tt L}x,{\tt R}y)={\tt C}(\hbox{\tt half\_sum}(x,y)).\]

Here is a second example where \(d_5\) and \(d_6\) are modified with
respect to \(d_2\) and \(d_4\).  We suppose \(d_1={\tt C}\),
\(d_2={\tt L}\), and \(d_3={\tt L}\).  In this case, the following
equalities hold:
\begin{eqnarray*}
{\hbox{\tt half\_sum}}({\tt CL}x,{\tt L}d_4y)
&=&\frac{\frac{x}{4}+\frac{1}{4}+\frac{d_4y}{2}}
{2}\\
&=&\frac{\frac{x}{2}+\frac{1}{2}+\frac{d_4y}{2}}
{2}\\
&=&{\tt L}(\hbox{\tt half\_sum}({\tt R}x,d_4y))
\end{eqnarray*}
In this case, it is not necessary to scrutinize \(d_4\), but the value
\(d_5\) is modified with respect to \(d_6\).  The equation can be re-written
as
\[{\hbox{\tt half\_sum}}({\tt CL}x,{\tt L}y)=
{\tt L}(\hbox{\tt half\_sum}({\tt R}x,y)).\]

If we had designed the algorithm to scrutinize two digits in each
input, there would be 81 cases, but since some cases can be handled
with a scrutiny of only the first digit in each input, or only one
digit in one of the inputs, the number of cases is brought down to 25
cases.  Moreover, the exact behavior of the algorithm in each case can
be found automatically with the help of proof search procedures.

\subsection{Automatic generation of function code}
We use the proof engine to actually construct the half-sum function by
making this program use its proof search facility to construct a
term with the right type, which should be
\begin{alltt}
  stream idigit -> stream idigit -> stream idigit
\end{alltt}
We have enough control on the proof search mechanism to express
that the value of this type that we want to construct should be
a co-recursive function and that it should analyze the first digit
of the two input streams.  This is simply done by stating that a 
case analysis on these arguments should be performed.  Doing a
case analysis on the first digit of the first input yields three
arguments, doing a case analysis on the first digit of the second argument
also gives three cases, so that there are at least nine cases to consider.
Some cases are easily solved directly, by simply finding an output
digit that makes the addition correct.  For instance, if the two
inputs are \(dx\) and \(dy\) (in other words, they have the same
initial digit) then the result should be \(d({\tt half\_sum}~x~y)\).
This because the formula holds:
\[\frac{(\frac{x}{2}+\frac{\alpha(d)}{2})+(\frac{y}{2}+\frac{\alpha(d)}{2})}
{2}=\frac{\frac{x+y}{2}}{2}+\frac{\alpha(d)}{2}.\]
When no output digit can be computed to make the formula work directly,
more information is gathered from the inputs by imposing more case analysis.
When this case analysis is performed, we look at possible values of
the second digit of one of the inputs and we decide if we have enough
information to decide what the first output digit should be.  This decision
is taken by performing some interval reasoning.  If two digits of one of
the inputs are fixed, then this input belongs in an interval of length
\(\frac{1}{4}\), adding this interval with an input for which only one digit
is known, this gives an interval of length \(\frac{3}{4}\), taking the half
of this gives an interval of length \(\frac{3}{8}\).  If the lower bound
of this interval is 0, \(\frac{1}{16}\), \(\frac{1}{8}\), \(\frac{1}{4}\),
\dots then we know what the output first digit can be, but if the
lower bound of this interval is \(\frac{3}{16}\) (this happens when one of
the inputs is {\tt LC}\(x\) and the other is {\tt C}\(y\)), then the upper
bound is \(\frac{9}{16}\) and we cannot conclude because
this interval may contains values lower than \(\frac{1}{4}\) (which should
not start with a {\tt C} or a {\tt R}) and values
larger than \(\frac{1}{2}\) (which should not start with a {\tt L}): for
these cases an extra case analysis on the second input is required.

When the first digit of the output is fixed, we still need to decide what
the first digit of the arguments to recursive calls will be.  This may
include a change with the initial second digit of the input.  This
difference is often related to the equivalence between {\tt LR} and {\tt CL}
prefixes on the one hand and between {\tt RL} and {\tt CR} on the other
hand.

For instance, if the first input has the form {\tt CL}\(x\) and the
second input has the form {\tt L}\(y\), the function can return
{\tt L}({\tt half\_sum}({\tt R}\(x\) \(y\))), because the half sum
of the two inputs is equivalent to the half sum of {\tt LR}\(x\) and
{\tt L}\(y\).

To determine the first digits of inputs in recursive calls, we must
first respect an important rules: variables that represent sub-streams
should appear behind the same number of digits in the input pattern and
in the output pattern.  This rules comes from the fact that the real number
represented by these variables is divided by 2 every time a digit is added
in front of it.  If we want the half-sum equation to be satisfied we must
make sure that the number of divisions by 2 is preserved between the inputs
and the output.  Thus, we can only prescribe the
first digit of one the arguments
to the co-recursive call of {\tt half\_sum} if the corresponding input had
two digits in the input pattern.

To determine what first digit can be used for the recursive call on an
argument, the proof search procedure compares the lower bound of the
output as prescribed by its first digit to the lower bound of possible results
that we can predict from the shape of the input patterns.
In the example, the lower bound of the output as prescribed by the first
digit {\tt L} is 0, the lower bound predicted from the half-sum of
{\tt CL}\(x\) and {\tt L}\(y\) is  \(\frac{1}{8}\).  The discrepancy must be
resolved by making sure that the sum of all the digits 
appearing as head of recursive call
arguments adds up to \(\frac{1}{2}\) (which does fit with a target
\(\frac{1}{8}\) since we are computing a half-sum and place the output's
first digit {\tt L} in front).  Here there is only one
digit available, and we can only choose its value to be {\tt R}.

Although the half-sum function is obtained by mechanical means to be
correct by construction, its type is only
\begin{alltt}
   stream idigit -> stream idigit -> stream idigit
\end{alltt}
This type does not express what the function does.
We need to add a theorem to state that it has the right behavior with
respect to the real numbers represented by the inputs and outputs.  The
statement has the following shape.
\begin{verbatim}
Theorem half_sum_correct :
  forall x y u v, represents x u -> represents y v ->
    represents (half_sum x y) ((u + v)/2).
\end{verbatim}
The proof of this statement relies on a proof by co-induction: we assume
that the statement is already satisfied
for any output stream that is a strict sub-stream of the current output and
we show that this is enough to prove the result for the current output.
The proof analyzes the behavior of the {\tt half\_sum}
function and explores all the 25 cases that were found at the
time the function was constructed.  In all cases, it is a simple matter
to verify the equality between the algebraic formula corresponding to
the half-sum of the inputs and the output pattern present in the
{\tt half\_sum} function, using a tactic named {\tt field} \cite{del.may.2001}
that solves equalities between rational expressions in a field;
a second statement that needs to be verified is
that the half-sum of the inputs does belong to [0,1] if the two inputs do,
this is easily done using a tactic named {\tt fourier} that solves affine
comparisons between real values with rational coefficients.

We believe that our definition technique can easily be reproduced
for different sets of digits or for other simple binary operations
like subtraction.

\subsection{Multiplication by 2}
We also need to provide a
function to  multiply the output of {\tt half\_sum} by 2.
This function is based on
the following remarks.

 \begin{itemize}
     \item The double of a number of the form \({\tt L}x\) is simply \(x\),
     \item The double of a number of the form \({\tt R}x\) is either 1
          or outside the interval [0,1],
     \item The double of a number of the form \({\tt C}x\) is a number
           of the form \({\tt R}x'\) where \(x'\) is the double of
           \(x\) (hence the algorithm exhibits a co-recursive call).
 \end{itemize}

Here is the formal definition:
\begin{verbatim}
Cofixpoint mult2 (v:stream idigit): stream idigit :=
  match v with L::x => x | C::x => R::mult2 x | R::x => one end.
\end{verbatim}

The correctness of this function is expressed with the following statement:
\begin{verbatim}
Theorem mult2_correct :  forall x u,
    0 <= u <= 1/2 -> represents x u -> represents (mult2 x)(2*u).
\end{verbatim}
Please note that this theorem explicitly states that the result value
is specified correctly only when the input is
smaller than \(\frac{1}{2}\).
\subsection{Subtraction}
In this section we discuss several approaches to subtraction.  One
first approach uses a few intermediary functions.  The first
intermediary function mimics the opposite function.  Of course the
opposite function cannot be defined from [0,1] to [0,1], but 
we can define the function
that maps \(x\) to \(1-x\).  Here is the general definition, where we
name the function {\tt minus\_aux}:
\begin{eqnarray*}
\hbox{\tt minus\_aux}({\tt L} (x)) &=& {\tt R}(\hbox{\tt minus\_aux}(x))\\
\hbox{\tt minus\_aux}({\tt C} (x)) &=& {\tt C}(\hbox{\tt minus\_aux}(x))\\
\hbox{\tt minus\_aux}({\tt R} (x)) &=& {\tt L}(\hbox{\tt minus\_aux}(x))
\end{eqnarray*}
These equations are justified through simple computations.  For instance,
the last equation is justified with the following reasoning steps:
\begin{eqnarray*}
\hbox{\tt minus\_aux}({\tt R} (x)) &=&
  1-(\frac{x}{2}+\frac{1}{2})\\
&=& \frac{1}{2}-\frac{x}{2}\\
&=& \frac{1}{2}(1-x)\\
&=& {\tt L}({\tt minus\_aux}(x))
\end{eqnarray*}

Combining {\tt minus\_aux} with addition, we can easily compute the
binary function that maps \(x\) and \(y\) to \(1+x-y\).  Of course, this function
returns a meaningful result only when \(x\) is smaller than \(y\).

Alternatively, we can combine 
{\tt minus\_aux} with {\tt half\_sum} to have a function that maps
\(x\) and \(y\) to 
\(\frac{1+x-y}{2}\).  Now, if we really want to have a subtraction, we can
remove the \(\frac{1}{2}\) offset.  We use another auxiliary function,
which we name
{\tt minus\_half} and is defined by the following equations:
\begin{eqnarray*}
\hbox{\tt minus\_half}({\tt R} x) &=& {\tt L} x\\
\hbox{\tt minus\_half}({\tt L} x) &=& {\tt zero}\\
\hbox{\tt minus\_half}({\tt C} x) &=& {\tt L}(\hbox{\tt minus\_half}(x))
\end{eqnarray*}
The first of these equations is trivial to justify.  The second is justified by the
fact that the only value inside \([0,\frac{1}{2}]\) for which \(x-\frac{1}{2}\) belongs
to \([0,1]\) is \(\frac{1}{2}\) and the result is 0 in this case.  The third equation
is justified by the following reasoning steps:
\begin{eqnarray*}
\hbox{\tt minus\_half}({\tt C} x) &=& \frac{x}{2}+\frac{1}{4} - \frac{1}{2}\\
&=&\frac{x}{2}-\frac{1}{4}\\
&=&\frac{x-\frac{1}{2}}{2}\\
&=&{\tt L}(\hbox{\tt minus\_half}(x))
\end{eqnarray*}
\section{Parameterized affine operations}
In this section, we study another approach to addition, with the encoding
of a more general function that computes affine formulas in two real
values with rational coefficients.   More precisely, we want to compute
the value
\[\frac{a}{a'}x+\frac{b}{b'}y+\frac{c}{c'}\]
When \(a\), \(b\), \(c\) are non-negative integers and \(a'\), \(b'\),
\(c'\)
are positive integers (\(a\), \(b\), \(c\) may be null, but not
the others), and \(x\) and \(y\) are real numbers, given as infinite
sequences of digits.

The interpretation of digits as affine functions (using our function
{\tt lift}) makes them a special case of what Edalat and Potts call {\em
Linear Fractional Transformations} \cite{edalat.98}.  They actually show that
a more general form of two argument transform can be programed on this
form of real number transformation, namely the computation of the
following transform, called a {\em M{\"o}bius transform}, where \(a\), \(b\), etc.
are integers:
\[\frac{axy +bx+cy+d}{exy+fx+gy+h}.\]
Studying only affine formulas correspond to restricting the general study
proposed by Edalat and Potts to the case of where \(e\), \(f\), and \(g\)
are 0.  A good reason to study separately the restricted case is that the
formal proofs stay inside the realm of proofs about equalities and
comparisons of affine formulas with rational coefficients, a realm for which
automatic proof tools exist at the time of writing this article, while
verifying the correctness of the general M{\"o}bius transform requires
incursions
into the realm of proofs about equalities and comparisons of polynomial
formulas, a domain for which proof procedures are only under development
\cite{MahboubiPottier2002,mclaughlin-harrison,Mahboubi06}.

\subsection{Main structure of the algorithm}
Choosing the digits of the result is based on the following remarks:
\begin{enumerate}
  \item Even without observing \(x\) and \(y\), we already know that
  they are between \(0\) and \(1\).  The result lies between
  the extrema
\[\frac{c}{c'}\qquad \hbox{and} \qquad \frac{ab'c'+a'bc'+abc'}{a'b'c'}.\]
\item\label{corec} if the extrema belong to the same basic interval,
  it is possible to produce a digit and perform a co-recursive call with a
  new affine formula.  In this sense, the output does not depend on reading
  more digits from the input and the algorithm can be described as a streaming
  algorithm in the sense of \cite{DBLP:conf/mpc/Gibbons04}.
\item \label{rec} If the extrema are badly placed, we cannot choose an
  interval associated to a digit that is sure to contain the result.
  In this case, we scrutinize \(x\) and \(y\) and observe their first
  digit.  As a result, we obtain a new estimate of the interval that
  may contain the result, whose size is half the previous size.  We
  can then perform a recursive call with a new affine formula.  In the
  long run, we are forced to get to a situation where the extrema are
  inside a basic interval and a co-recursive call can be performed.
  In fact, this
  condition is guaranteed as soon as the distance between extrema is
  shorter than \(1/4\).
\end{enumerate}

Let us study two examples.  In the first example, suppose that the
property \(\frac{c}{c'}\ge 1/2\) holds.  We know that the result is larger
than \(1/2\) and we can produce a {\tt R} digit.  The following
computation takes place:
\begin{eqnarray*}
\frac{a}{a'}x+\frac{b}{b'}y+\frac{c}{c'}&=&
 {\tt R}(2\times(\frac{a}{a'}x+\frac{b}{b'}y+\frac{c}{c'})-1)\\
&=&R(\frac{2a}{a'}x+\frac{2b}{b'}y+\frac{2c-c'}{c'})
\end{eqnarray*}
There is a recursive call with a new affine formula, where all the
coefficient are positive integers or non-negative integers as required.

In a second example, suppose that the properties \(x={\tt L}x'\) and
\(y={\tt R}y'\) hold.  The following computation can take place:
\begin{eqnarray*}
\frac{a}{a'}x+\frac{b}{b'}y+\frac{c}{c'}&=&
 (\frac{a}{a'}\frac{x'}{2})+(\frac{b}{b'}\frac{y'+1}{2})+\frac{c}{c'}\\
&=&\frac{a}{2a'}x'+\frac{b}{2b'}y'+\frac{bc'+2b'c}{2b'c'}
\end{eqnarray*}
Here again, we can have a recursive call with a new affine formula,
no digit has been produced (therefore the recursive call cannot be a
co-recursive call) but the distance between the extrema in the new
formula is \(a/2a'+b/2b'\), the half of \(a/a'+b/b'\), which was the
distance between extrema for the initial affine formula.

\subsection{Formal details for affine formulas}
When providing the formal description of the recursive algorithm for
the computation of affine formulas, we need to pay attention to
the following two important aspects:
\begin{enumerate}
  \item The function needs to be partial, because we must ensure the
sign conditions on the coefficients of the affine formula,
  \item Not all recursive calls are acceptable co-recursive calls, because
    recursive calls after the consumption of digits from the two input
    streams are associated to no production of an output digit.
\end{enumerate}
We define a record type named {\tt affine\_data} that collects the eight
elements of an affine formula and a predicate named 
{\tt positive\_coefficients} that states that the
coefficients satisfy the sign conditions.

The computation is then represented by a main function called
{\tt axbyc} to compute the affine formula.  This function has a dependent
type: it takes as first argument an affine formula and as second argument
a proof that its coefficients satisfy the sign conditions.
\begin{verbatim}
axbyc: forall x: affine_data, positive_coefficients x -> stream idigit.
\end{verbatim}
This function is defined as a co-recursive function.  The constraints
on recursive programming impose that this function can only perform the
recursive calls that are associated to the production of a digit in the
output (phase~\ref{corec} in the previous section).  We need to define an
auxiliary function, not a co-recursive
one, which takes charge of the recursive calls that are only associated
to the consumption of digits from the input streams (phase~\ref{rec} in
the previous section).

The auxiliary function is named {\tt axbyc\_rec}.  It takes as arguments
an affine formula and a proof that it satisfies the predicate
{\tt positive\_coefficients}.  It returns an equivalent affine formula,
one for which the decision of producing the next output digit
can be taken.  This is represented by the fact that output of this
function is in a type with three constructors,
called {\tt caseR}, {\tt caseL}, or {\tt caseC}.  Each constructor 
contains as first field the new affine formula, as second field a proof
that this new formula satisfies the sign conditions.  The next field
(for the constructors {\tt caseR} and {\tt caseL})
or the next two fields (for the constructor {\tt caseC}) express that the
right interval conditions are satisfied to output a digit, The last field
is a proof that the new affine formula is equivalent to the initial one.

The recursive structure of the function {\tt axbyc\_rec} is based on
well-founded recursion.  More precisely, it relies on the fact that
the distance between extrema decreases at each recursive step.  This
can be translated into an integer formula that decreases while remaining
positive.  When this integer formula is 0, we can prove that one of the
interval conditions to output a digit is necessarily satisfied.

Two other collections of auxiliary function perform the elementary operations.
Functions named {\tt prod\_R}, {\tt prod\_L}, and {\tt prod\_C} perform
the coefficients transformations that should be performed after producing
an output digit.  For instance {\tt prod\_R} maps the affine formula
\[\frac{a}{a'}x+\frac{b}{b'}y+\frac{c}{c'}\]
to the formula 
\[\frac{2a}{a'}x+\frac{2b}{b'}y+\frac{2c-c'}{c'}\]
as we already justified in the first example of the previous section.

The lemmas named {\tt A.prod\_R\_pos}, {\tt A.prod\_L\_pos}, and 
{\tt A.prod\_C\_pos} ensure that the functions {\tt prod\_R}, {\tt prod\_L}
and {\tt prod\_C} preserve the sign conditions on the coefficients, 
respectively.  These lemmas rely on the interval conditions produced
by {\tt axbyc\_rec}.  For instance, for {\tt A.prod\_R\_pos} the extra interval
condition is \(c'\leq 2c\).  In our description of the co-recursive
function, this information is transferred from {\tt axbyc\_rec} to
{\tt A.prod\_R\_pos} with the help of a variable named {\tt Hc}.

With these auxiliary functions, the main function can be given a simple
structure:
\begin{verbatim}
CoFixpoint axbyc (x:affine_data) 
   (h:positive_coefficients x):stream idigit :=
  match axbyc_rec x h with
    caseR y Hpos Hc _ =>
     R::(axbyc (prod_R y) (A.prod_R_pos y Hpos Hc))
  | caseL y Hpos _ _ =>
     L::(axbyc (prod_L y) (A.prod_L_pos y Hpos))
  | caseC y Hpos H1 H2 _ =>
     C::(axbyc (prod_C y) (A.prod_C_pos y Hpos H2))
  end.
\end{verbatim}

With the help of the function {\tt real\_value} we can also define
a function {\tt af\_real\_value} that maps any affine formula
represented by an element of {\tt affine\_data} to the real number it
represents.  This function is used to express the correctness of our
algorithm to compute the affine formula:
\begin{verbatim}
axbyc_correct:
  forall x, forall H :positive_coefficients x,
   0 <= af_real_value x <= 1 ->
   real_value (axbyc x H) = af_real_value x.
\end{verbatim}
This proof is based on a lemma that is proved by co-induction:
\begin{verbatim}
axbyc_correct_aux :
  forall x:affine_data, forall H :positive_coefficients x,
  0 <= (af_real_value x) <= 1 ->
   represents (axbyc x H) (af_real_value x).
\end{verbatim}
This algorithm is interesting because it provides ways to
add two real numbers, to multiply them by rational numbers, and to encode
rational numbers as real numbers.  Having formalized this algorithm may
make the direct implementation of addition described earlier seem useless.
It is not useless, because the direct implementation of addition makes no
use of dependent types, proof arguments, or well-founded recursion.  As a
result, the direct addition can be executed directly inside the theorem
prover using its internal reduction mechanism, while the affine formula
computation can only be executed outside the theorem prover as extracted
code.  Algorithms that can be reduced inside the proof system may play
a role in reflection-based proof procedures \cite{Boutin97b}.
\section{Computing series}
A series is an infinite sum of values.  Knowing how to compute series
can help in computing famous constants like \(e\) (Euler's number) or
\(\pi\) and
to implement the multiplication of two real numbers.
\subsection{Main structure of the algorithm}
We consider the problem of computing values of the form
\(\sum_{i=0}^{\infty} a_i\) where the \(a_i\) terms are real numbers.

Studying series is very close to
studying converging sequences, since it is enough to consider the
sequence \(u_n=\sum_{i=0}^n a_i\).  Each element of the sequence can
then be computed as a finite combination of additions.

Computing the first \(p\) digits of the limit
means computing the limit up to a precision of \(2^{-p}\).  If we know that a
given element \(u_n\) is closer than \(2^{-(p+1)}\) to the
limit and if we can compute \(u_n\) to a precision better than
\(2^{-{p+1}}\), then we are able to compute the limit to the
required precision.  In other words, if we know that
\(|\sum_{i=n}^{\infty} a_i|\) is smaller than \(2^{-(p+1)}\), and we
know a value \(v\) whose distance to \(\sum_{i=0}^n a_i\) is less
than \(2^{-(p+1)}\), then we also know that the distance of \(v\) to
is less than \(\sum_{i=0}^{\infty} a_i\).   In fact, we need to use
slightly stronger precisions, because an interval of length \(2^{-p}\) rarely
fits in one of the intervals representable by finite sequences of digits.
Still, this approach shows that we can avoid considering the whole infinite sum
to produce the first digit of the output.

We restrict our study to series whose convergence is described
constructively by a function \(\mu\) that satisfies the following properties:
\[\forall m.\quad n\le m \Rightarrow |\sum_{i=m}^\infty a_i| < \mu(n)
\qquad \lim_{n\rightarrow\infty}\mu(n)=0\]  
We actually formalize the computation of a function \(f\) that has the
following informal specification:
\[ f(x,y,n) = x + y\times\sum_{i=n}^\infty a_i,\]
where \(x\) is a real number, \(y\) is a rational number,
and \(n\) is a natural number. Intuitively, \(y\) represents the inverse of
the precision that is reached in the computation (\(y=2^p\)).
If we know that \(y\times\mu(n)\leq \frac{1}{16}\) and we know \(x\)
to a precision of three digits then we are able
 to choose the first digit \(d\) of  \(x+y\sum_{i=n}^\infty a_i\).
We can then perform the following computation
\[f(x,y,n)=d\,f(2x-\alpha(d),2y,n),\]
In most cases, we also have \(x=dx'\) for some \(x'\), and
the value \(2x-\alpha(d)\) is simply
represented by \(x'\).  If \(y\times\mu(n)> \frac{1}{16}\), we compute
a new value \(\phi(y,n)\) such that \(y\times\mu(\phi(y,n))\leq\frac{1}{16}\).
This value is bound to exist because \(\mu\) converges to 0 at infinity.
We can then proceed with the following step.
\[f(x,y,n)=f(x+y\times\sum_{i=n}^{\phi(y,n)-1}a_i,y,\phi(y,n))\]

In practice we first compute \(\phi(y,n)\), then we compute
\(v=x+y\times \sum_{i=n}^{\phi(y,n)-1} a_i\) by repeated
binary additions.  Let us assume that
\(\rho\) is defined as \(y\times \sum_{i=\phi(y,n)}^\infty a_i\).
 The number we want to compute is \(v+\rho\) and we know that
\(|\rho|\leq \frac{1}{16}\). We then perform the following case
analysis:
\begin{enumerate}
\item If \(v\) has the form \({\tt R}v'\), and \(v'\) has the form
{\tt R}\(v''\), \({\tt C}v''\), {\tt  LC\(v''\)}, or {\tt LR\(v''\)}, 
we can deduce that \(v\geq 9/16\), therefore
\(v+\rho \geq \frac{1}{2}\) and the first
digit of the result can be {\tt R}.  The 
result is \({\tt R}(f(v',2y,\phi(y,n)))\).
\item If \(v\) has the form {\tt C}\(v'\), where \(v'\) has the form
{\tt C}\(v''\), {\tt LC}\(v''\), {\tt LR}\(v''\), {\tt RL}\(v''\), or
{\tt RC}\(v''\), then we are certain that
\(\frac{5}{16}\leq v\leq \frac{11}{16}\), therefore \(\frac{1}{4}\leq v+\rho
\leq \frac{3}{4}\), the result is \({\tt C}(f(v',2y,\phi(y,n)))\).
\item If \(v\) has the form {\tt L}\(v'\), where \(v'\) has the form
{\tt L}\(v''\), {\tt C}\(v''\), {\tt RL}\(v''\), {\tt RC}\(v''\), then we
are certain that \(v\leq \frac{7}{16}\) and \(v+\rho\leq \frac{1}{2}\)
the result is \({\tt L}(f(v',2y,\phi(y,n)))\).
\item if \(v\) has the form {\tt RLL}\(v''\), then \(v\) could also be
  represented using {\tt CRL}\(v''\) and this case is already
  considered above.  The same goes for the cases {\tt LRR}, {\tt CLL}, and
  {\tt CRR}, using {\tt CLR}, {\tt LRL}, and {\tt RLR} as respective
alternatives.
\end{enumerate}
The number \(\phi(y,n)\) is chosen so that
 \(y\times\mu(\phi(y,n))\leq \frac{1}{16}\) because \(\frac{1}{16}\)
is the shortest distance between the bounds of the intervals
for {\tt CLC} and {\tt C}, {\tt RLC} and {\tt R}, or {\tt LRC} and {\tt L}.
Computing \(\phi(y,n)\) and  \(\sum_{i=n}^{\phi(y,n)-1}a_i\)
depends on the series being studied.  Because recursive calls to
\(f\) always have \(2y\) and \(\phi(y,n)\) as arguments, we can also
assume
that the invariant \(y\times\mu(n)<\frac{1}{8}\) is maintained through
recursive calls.
\subsection{Series with positive terms}
When we know that the \(a_i\) terms are all positive, we do not need
to use \(\frac{1}{16}\) to bound the infinite remainder of the series.
The computation technique can be simplified.

We first compute \(\phi(y,n)\) so that \(y\times\mu(\phi(y,n))
 \leq \frac{1}{8}\)
and \(v=x+\sum_{i=n}^{\phi(y,n)-1}\). 
In what follows, let \(\rho\) be defined as 
\(y \times\sum_{i=\phi(y,n)}^\infty a_i\); we know \(|\rho|\leq \frac{1}{8}\).
We perform the following case analysis:
\begin{enumerate}
  \item if \(v\) has the form \({\tt R} v'\), we are sure that the
  result is larger than or equal to \(1/2\), the result is \({\tt
  R}(f(v',2y,\phi(y,n)))\).
  \item if \(v\) has the form \({\tt C} v'\) but not \({\tt CR} v'\),
   we can deduce \(v \in [\frac{1}{4},\frac{5}{8}]\) and
 \(v+\rho\in[\frac{1}{4},\frac{3}{4}]\), the result is
   \({\tt C}(f(v',2y,\phi(y,n)))\).
\item if \(v\) has the form \({\tt L}v'\), but not
\({\tt LR}v'\), we can deduce \(v\in [0,\frac{3}{8}]\) and
 \(v+\rho\in [0,\frac{1}{2}]\).
The result is \({\tt L}(f(v',2y,\phi(y,n)))\).
\item if \(v\) has the form \({\tt CR}v''\) or \({\tt LR}v''\), we can use
 the equivalences with \({\tt RL}v''\) and \({\tt CL}v''\),
 respectively, to switch to one of the previous cases.
\end{enumerate}
For positive series, there is also an invariant: \(y\times\mu(n)\) is
always smaller than \(\frac{1}{4}\).  This invariant plays a role
in the correctness proofs.
\subsection{Formal details of positive series}
We programmed the general treatment of positive series in a
higher-order function that is easy to re-use from one series to the
other.
\begin{verbatim}
Definition series_body (A:Set) 
  (f: stream idigit -> A -> stream idigit) 
  (x: stream idigit) (a:A): stream idigit :=
 let (d,x') := v in
 match d with
   R =>  R::f v' a
 | L => match v' with R::v'' => C::f (L::v'') a | _ => L::f v' a end
 | C => match v' with R::v'' => R::f (L::v'') a | _ => C::f v' a end
  end.
\end{verbatim}
The type {\tt A} should make it possible to compute
the values
\(y\) and \(n\) used in our informal presentation.  This will be made
clearer in the next examples.
\subsection{Application to computing $e$}
The number \(e\) is defined by the formula
\[e=\sum_{k=0}^\infty \frac{1}{k!}.\]
Of course, this number is larger than 2, but we only want to compute
its fractional part, so that we actually compute
\(\sum_{k=2}^\infty \frac{1}{k!}\).
The following properties are easy to obtain, by remembering that
\(n!n^{k-n}<k!\) for every \(k \ge n\), therefore
\[0 < \sum_{k=n}^{\infty} \frac{1}{k!} < \frac{1}{(n-1)!(n-1)}.\]
For this series, we choose \(\mu(n)\) to be the value
\(\frac{1}{(n-1)!(n-1)}\).  As soon as
\(2<n\), we have \(\mu(n+1)<\frac{\mu(n)}{2}\).  This implies the
following property:
\[\forall n, y, 0 < y \wedge 2 < n \wedge y\times\mu(n) < \frac{1}{4}
 \Rightarrow \phi(y,n) \leq n+1.\]
Thus, it is never necessary to absorb more than one term from the
infinite sum into \(x\) at each co-recursive call.

the type {\tt A} that appears in our use of {\tt series\_body} is a
triple type.  The triple given as argument has the form 
\((y,n,\theta)\), where \(\theta\) is the precomputed value 
\(\theta=(n-1)!\).  This invariant is maintained through recursive calls
so that factorials are not recomputed from scratch each time.  Here the
\(\mu\) function is given by the formula
\[\mu(n)=\frac{1}{(n-1)!(n-1)}=\frac{1}{\theta\times(n-1)},\]
and computing 
\(\phi(y,n)\) is easy, because we know that this value is always
\(n\) or \(n+1\).  To decide whether \(\phi(y,n)=n\), 
we simply need to compare
\(y\times\frac{1}{\theta(n-1)}\) with \(\frac{1}{8}\),
in other words to compare \(8y\) with \(\mu'=\frac{1}{\mu}=\theta(n-1)\).
In the following code, {\tt rat\_to\_stream} builds the infinite sequence
of digits for a rational number given by its numerator and denominator.
\begin{verbatim}
CoFixpoint e_series (x:stream idigit)(s :Z*Z*Z) :stream idigit :=
  let (aux, theta) := s in let (y, n) := aux in let mu' := theta*(n-1) in
  let (v, phi, theta') := 
   if Z_le_gt_dec (8*y) mu' then
      mk_triple x n theta
   else
      let theta' := mu'+theta in
      mk_triple (x+(rat_to_stream y theta'))(n+1) theta' in 
   series_body  _ e_series v (2*y, phi, theta').
\end{verbatim}

To express that this function computes correctly the series, we
rely on a predicate {\tt infinite\_sum} which takes a function \(f\)
from {\tt Z} to {\tt R} and a value \(v\) in {\tt R} as arguments
and means {\em \(\sum_{i=0}^\infty f(i)\) converges and the limit is \(v\).}
The correctness statement has the following shape:

\begin{verbatim}
Theorem e_correct1 :
 forall v vr r y n, 
   4 * y <= fact(n-1) * (n-1) -> 2 <= n -> 1 <= y ->
   represents v vr ->
   infinite_sum (fun i => 1/IZR(fact (i+n))) r ->
   vr + (IZR y)*r <= 1 ->
   represents (e_series v (y, n, fact(n-1))) (vr+(IZR y)*r).
\end{verbatim}
This statement is obfuscated by the simultaneous use of two
types of numbers and the function {\tt IZR} is the natural injection
of integers into the type of real numbers.
In this statement the formula \verb+fact(n-1)*(n-1)+ represents the
inverse of \(\mu(n)\) and the formula
\verb"4 * y <= fact(n-1) * (n-1)"
 corresponds to the invariant of the series.
  Note that this theorem
expresses that the series is correctly computed only if the series 
really converges to a value that is smaller than or equal to 1.  The
proof that the series converges has to be done independently.

We initialize the recursive computation with \(x=\frac{1}{2}\),
\(y=1\) and \(n=3\), so that the invariant on \(y\times\mu(n)\) is
satisfied.
\begin{verbatim}
Definition number_e_minus2: stream idigit :=
  e_series (rat_to_stream 1 2+rat_to_stream 1 6) (1, 4, 6).
\end{verbatim}
The correctness theorem then makes it possible to obtain the
following statement:
\begin{verbatim}
Theorem e_correct : 
  infinite_sum (fun i => 1/IZR(fact(i+2)))(real_value number_e_minus2).
\end{verbatim}
This statement really means \(\sum_{i=2}^\infty \frac{1}{i!}={\tt number\_e\_minus2} \).

The value {\tt number\_e\_minus2} can then be used to construct
rational approximations of \(e-2\), using the {\tt bounds} function.
  Actually, given \(n\)
we compute \((a,b,k)\) so that
\[\frac{a}{2^k} \leq e - 2 \leq \frac{b}{2^k}\qquad \frac{b}{2^k} -
\frac{a}{2^k} = \frac{1}{2^n}.\]
For \(n=320\), this computation takes approximately a minute with the
standard version of 
\coq{}\footnote{Coq version {\tt 8.0pl2}, Intel Pentium(R) M 1700Mhz.}.
These functions can be used in proof procedures
to compute approximations
of \(e\).
The code can also be extracted both to Ocaml and to Haskell.  Running
the Ocaml extracted code, we 
can compute 2000 redundant digits of \(e-2\) in about a minute.
\subsection{Multiplication as a special case of series}
When \(u\) is the infinite sequence \(d_1d_2\dots\) and then \(uu'\)
is a series:
\[uu'=\sum_{i=1}^{\infty}\frac{\alpha(d_i)u'}{2^i}.\]
This is a series where all terms are positive.  Moreover, two
simplifications can be made with respect to the general approach.
First, while \(y\) is multiplied by 2 at every recursive call, \(a_i\)
contains a divisor that is also multiplied by \(2\), so that the two
multiplications by 2 cancel out.  Second, it is reasonable to simply
consume one element from the infinite sum at each recursive call, without
scrutinizing the value of \(u'\).  If this approach is
followed, the argument \(y\) is not necessary anymore: only the digits
\(d_i\) and \(u'\) are needed.  We can re-use the general function
{\tt series\_body} in the following manner:
\begin{alltt}
Definition sum_mult_d (d:idigit) (u v:stream idigit) :=
  match d with L => u | C => u+L::L::v | R => u+L::v end.

CoFixpoint mult_a (x:stream idigit)(p:stream idigit*stream idigit) 
  : stream idigit :=
 let (u,v) := p in
 match u with d::w => series_body _ mult_a (sum_mult_d d x v)(w,v) end.
\end{alltt}
The function {\tt mult\_a \(x\) \((u,u')\)} computes \(x+uu'\) only when
\(uu' < \frac{1}{4}\) (here again, we see the invariant of the general
approach).  To obtain multiplication for the general case, we divide
the second operand by \(4\) before the multiplication and we multiply
the result by \(4\).
Here is a naive implementation:
\begin{verbatim}
Definition mult (x y:stream idigit): stream idigit :=
  mult2(mult2(mult_a zero (x,L::L::y))).
\end{verbatim}
The following theorem can then proved and verified formally:
\begin{verbatim}
mult_correct
   : forall (x y: stream idigit) (vx vy: Rdefinitions.R),
     represents x vx -> represents y vy -> represents (x*y)(vx*vy)
\end{verbatim}
In this statement the notation {\tt x * y} represents our
multiplication as an operation on infinite digit streams, while the
notation
{\tt vx * vy} represents the multiplication of real numbers, as they
are axiomatized in the {\coq} system.

It turns out
our approach of multiplication, based on series, yields an
implementation of multiplication that is very close to the
implementation in \cite{ciaffaglione.00}.

We can also try this multiplication directly inside \coq{}, for
instance, we can compute \((e-2)^2\).  With the current standard
version of {\coq} no effort is made to
exploit possible sharing in the lazy computation of values, so that
the same value may be computed several times.  For this reason, we
cannot compute this number to a high precision as easily as for
\(e-2\).  For example, our few experiments showed that it takes
approximately 10 seconds to compute an approximation with an accuracy
of \(\frac{1}{2^{30}}\) digits and a minute
to compute the approximation with an accuracy of \(\frac{1}{2^{50}}\).

\section{Conclusion}
The work described in this paper is available on the Internet at the
following address:
\begin{verbatim}
  http://hal.inria.fr/inria-00001171
\end{verbatim}

This is both an extension and a departure
from the work of Ciaffaglione and di~Gianantonio.  We chose to work
with an extra digit that is interpreted as \(\frac{1}{2}\) instead of
\(-1\).  We believe algorithms are easier to design in this approach
and easier to prove correct, but this approach is less well adapted to
expanding to the full real line.

We believe the choice to rely on an existing axiomatization of real numbers
was instrumental in making the proofs in this experiment quicker to
obtain.  In particular, we relied on a collection of automatic decision
procedures for equalities between polynomials and sets of in-equalities
between affine formulas.  This axiomatization relies on classical mathematics,
and we don't know in which respect the decision procedures rely on these
these classical axioms and in which respect they could be re-implemented
in constructive mathematics.  Admittedly, we only provide algorithms to
compute representations of real values that are constructively definable.
We believe that our experiment should be reproducible in the context of
constructive mathematics, but it concentrated on the presentation of real
numbers that provided the most complete theory of functions.  The question
of constructive or non-constructive mathematics was secondary in this
experiment.

It is characteristic that the definition of series appears to be more
basic than multiplication, but this is simply due to the fact that
the digit-based representation of numbers already is naturally interpretable
as a series and this structure is preserved by multiplication thanks to
distributivity.

Now that we have a multiplication for our representation of real
numbers, we can consider the task of implementing other functions,
like division and analytic functions.  For division, we expect
to use a method of range reduction: to compute \(\frac{x}{y}\) we
should compute \(\frac{x}{2^ky}\) or \(\frac{x}{y}-k\) where \(k\)
should be chosen so that the result is inside \([0,1]\) but finding
the right value for \(k\) is only possible when we have a constructive
way to prove that \(y\) is non zero.

An alternative approach to division is provided if we generalize our
approach to affine formulas to consider M{\"o}bius transforms (already
studied in \cite{edalat.98}):
\[(x,y)\mapsto \frac{axy + bx + cy +d}{exy+fx+gy+h}.\]
However, we suspect that the proofs of correctness for these
transformations are less easy to automatize, because they do not rely
only on affine formulas (which are easily treated with the
Fourier-Motzkin decision procedure.

One of the interesting features of this experiment is that some series
can be computed directly inside the theorem prover.  This is an important
feature, if a proof requires that we produce an accurate approximation of 
a series value.

Concerning the computation of mathematical constants, we already have all the
ingredients to compute \(\pi\) using a Machin formula (we used
the formula \(\frac{\pi}{4}=\arctan(\frac{1}{2})+\arctan(\frac{1}{3})\) which
can easily be proved and computed as the sum of two series).

Several questions will be studied in the future.  First, the choice of
a digit set is arbitrary.  We already experimented we a digit set that
contains only two digits, corresponding to intervals \([0,\frac{2}{3}]\)
and \([\frac{1}{3},1]\); although some functions seem to be simpler (because
there are less cases to consider, for instance when considering affine
formulas) other problems arise because equalities between patterns do not
exist for short patterns (thus we cannot as easily mimic the equality
\({\tt CL} = {\tt LR}\)).  di~Gianantonio also studied a two-digit
representation and he proposes to work with 
intervals whose length is based on the golden number
\cite{diGianantonioGolden96}.  However, the price
to pay is that we now need to solve a polynomial system of degree 2 to
determine whether a given value belongs to one of the basic intervals, again
formal proofs in this setting are made more complex because we are not
in the real of affine formulas anymore.

In the long run, we wish to choose digit sets that are closer to the 
bound integers that are usually handled in computers, so that regular
integer addition, subtraction, multiplication, and comparison or even
bitwise operations can be used to establish the basic
relations between various digits.

The second important question is related to the efficiency of co-recursive
computation inside the theorem prover.  While recent evolutions of the \coq{}
system brought drastic improvements in the computation of recursive functions,
it is not certain that the co-recursive question is as well treated.  In
particular, lazy computation is needed to achieve reasonable speeds, while
the current version of {\coq} may only implement call-by-name.  This
question is particularly difficult because the reduction mechanism also needs
to retain the property of strong normalization for non-closed terms.

Eventually, we hope to achieve the development of a reasonably efficient
and formally verified library for mathematical computation.  We believe this
will be a stepping stone for more ambitious projects like the Flyspeck
project\cite{Hales00,hales-flyspeck}.

\bibliographystyle{apalike}
\bibliography{a,aspecific}
\end{document}